\documentclass[english,prl, twocolumn, longbibliography]{revtex4-1}
\usepackage[T1]{fontenc}
\usepackage[latin9]{inputenc}
\usepackage{array}
\usepackage{bm}
\usepackage{amsmath}
\usepackage{amssymb}
\usepackage{graphicx}

\makeatletter

\providecommand{\tabularnewline}{\\}

\usepackage{amsmath}
\usepackage{subfigure}
\usepackage{graphicx, mathtools}
\usepackage[colorlinks=true,linkcolor=MidnightBlue,urlcolor=MidnightBlue,citecolor=MidnightBlue,anchorcolor=MidnightBlue]{hyperref}\usepackage[dvipsnames]{xcolor}

\makeatother

\usepackage{babel}
\begin{document}

\title{Hydrodynamically interrupted droplet growth in scalar active matter}

\author{Rajesh Singh }

\affiliation{DAMTP, Centre for Mathematical Sciences, University of Cambridge,
Wilberforce Road, Cambridge CB3 0WA, UK}

\author{M. E. Cates}

\affiliation{DAMTP, Centre for Mathematical Sciences, University of Cambridge,
Wilberforce Road, Cambridge CB3 0WA, UK}
\begin{abstract}
Suspensions of spherical active particles often show microphase separation.
At a continuum level, coupling their scalar density to fluid flow,
there are two distinct explanations. Each involves an effective interfacial
tension: the first mechanical (causing flow) and the second diffusive
(causing Ostwald ripening). Here we show how the negative mechanical
tension of contractile swimmers creates, via a self-shearing instability,
a steady-state life cycle of droplet growth interrupted by division
whose scaling behavior we predict. When the diffusive tension is also
negative, this is replaced by an arrested regime (mechanistically
distinct, but with similar scaling) where division of small droplets
is prevented by reverse Ostwald ripening. 
\end{abstract}
\maketitle
Active matter continuously dissipates energy locally to perform mechanical
work. In consequence, the dynamical equations of coarse-grained variables,
such as particle density, break time-reversal symmetry. Examples include
suspensions of spherical autophoretic colloids, which self-propel
due to self-generated chemical gradients at their surfaces \cite{ebbens2010pursuit}.
Experiments on several such suspensions have observed activity-induced
phase separation that arrests at a mesoscopic scale \cite{palacci2013living,theurkauff2012dynamic,buttinoni2013DynamicClustering,ginot2018aggregation}.
A generic understanding of such nonequilibrium microphase separations
can be sought at the level of continuum equations for a diffusive
scalar concentration field, coupled to incompressible fluid flow.
Such an approach is complementary to more detailed mechanistic modelling,
in which particle motion and/or chemical fields are modelled explicitly
\cite{liebchen2015clustering,liebchen2017phoretic,alarcon2017morphology,sahaPRE2014,agudoPSPRL2019}.
By sacrificing detail, the resulting `active field theory' allows
maximal transfer of ideas and methods from equilibrium statistical
mechanics. Each distinct mode of activity can be modelled in a minimal
fashion, allowing the competition between them to be studied. This
knowledge base can inform the design of novel functional materials
and devices with tunable properties \cite{stenhammar2016light}.

Bulk active phase separation can arise through attractive interactions,
as in the passive case \cite{hohenberg1977theory}, or, even for repulsive
interactions, be motility-induced \cite{tailleur2008statistical,cates2015}.
At continuum level, when there is no orientational order in bulk,
the only order parameter required for the particles is their scalar
concentration $\phi({\bm{r}},t)$. For `wet' systems, with a momentum-conserving
solvent rather than a frictional substrate, this is coupled to a fluid
velocity field ${\bm{v}}({\bm{r}},t)$ \cite{marchetti2013}. Operationally,
the field theory of active scalar phase separation starts from the
long-studied passive case, whose stochastic equations of motion are
constructed phenomenologically, respecting symmetries and conservation
laws, truncated at some consistent order in the fields and their gradients
\cite{hohenberg1977theory}. If $\phi$ is measured relative to the
critical point for phase separation, this gives to leading order a
symmetric free energy functional ${\cal F}[\phi]={\cal F}[-\phi]$.
The outcome is Model B (dry) or Model H (wet) \cite{hohenberg1977theory,cates2018theories}.

To make these theories active, we add a small number of leading-order
terms, each of which breaks time-reversal symmetry via a distinct
channel. The first adds to the chemical potential $\delta{\cal F}/\delta\phi$
a piece that is not the derivative of any free energy functional ${\cal F}$
(this is the $\lambda$ term in \eqref{eq:orderPara-1} below). The
deterministic diffusive current then breaks detailed balance but remains
curl-free. This channel is known to alter phase boundaries, but cannot
arrest phase separation \cite{wittkowski2014scalar}. A second channel
(the $\zeta$ term in \eqref{eq:orderPara-1} below) enters at the
same order, but in the diffusive current directly. This can arrest
phase separation, by creating an effectively negative interfacial
tension in the diffusive sector \cite{tjhung2018cluster}, throwing
the process of Ostwald ripening, whereby large droplets grow at the
expense of small ones, into reverse. These two channels are, for dry
systems, captured in Active Model B+ \cite{tjhung2018cluster}.

The third active channel, present for wet systems only, is the mechanical
stress arising from self-propulsion. This is a bulk stress in systems
with orientational order (where it leads to bacterial turbulence \cite{marchetti2013}),
but for a scalar it is quadratic in $\boldsymbol{\nabla}\phi$ (the
$\tilde{\kappa}-\kappa$ term in \eqref{eq:activeStress} below),
just like the passive thermodynamic stress for Model H \cite{bray1994theory}.
It likewise drives fluid motion via interfacial curvature.

So far, the resulting `Active Model H' has been explored only in the
absence of the other two active channels, without noise, and for systems
at the critical density, $\phi_{0}\equiv\langle\phi\rangle=0$ \cite{tiribocchi2015active}.
Under these conditions, activity can arrest separation, giving a dynamically
fluctuating bicontinuous state. The effect of active stresses on droplet
states, as might describe the cluster phases seen experimentally \cite{ebbens2010pursuit,palacci2013living,theurkauff2012dynamic,buttinoni2013DynamicClustering},
remains unclear. Notably though, in passive systems, the interfacial
stress rapidly becomes unimportant on moving away from bicontinuity,
since well separated droplets recover spherical symmetry which precludes
incompressible fluid flow. Wet and dry dilute passive droplets then
behave similarly \cite{cates2018theories}.

In this Letter we first extend the study of Active Model H, with noise,
to off-critical quenches ($\phi_{0}\neq0$), where droplets or bubbles
arise, and there address the competition between mechanical ($\tilde{\kappa}$)
and diffusive ($\lambda,\zeta$) activity channels. We find that,
contrary to the passive case, the mechanical stress plays a crucial
role in droplet (or bubble) evolution; when sufficiently contractile,
it can halt phase separation, causing large droplets to split in smaller
ones. This balances the diffusive droplet growth due to Ostwald ripening,
giving a steady state with a distinctive droplet life cycle. This
represents `interrupted' rather than `arrested' phase separation,
because the steady state is highly dynamic, and continues unchanged
after all noise is switched off. This contrasts with the microphase
separation that results from reverse Ostwald ripening, where switching
off noise leads to a fully arrested, static assembly of monodisperse
droplets \cite{tjhung2018cluster}.

Crucially, the droplet life-cycle just described for the case of hydrodynamic
interruption requires splitting to be balanced by {\em forward}
Ostwald ripening, which sustains a stationary droplet number. In contrast,
we find that when the mechanical and diffusive activity {\em both}
favor microphase separation, the final steady state is arrested, not
interrupted. Despite this, the final droplet size depends on the active
stress parameter $\tilde{\kappa}$, instead of being fixed by either
the noise level or the initial condition, as happens for the dry case
of Active Model B+ \cite{tjhung2018cluster}. Our work thus exposes
a subtle interplay between different channels in the physics of active
microphase separation.
\begin{figure*}[t]
\centering\includegraphics[width=0.94\textwidth]{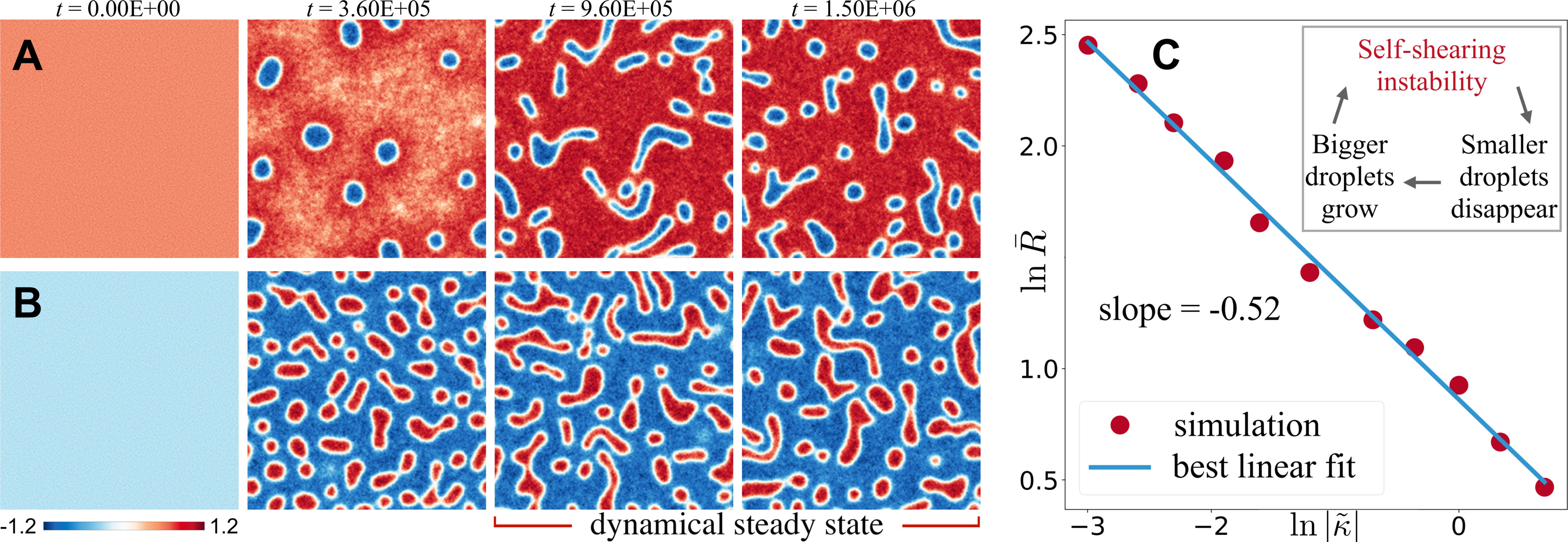}\caption{The self-shearing instability interrupts phase ordering in two dimensions
with contractile active stress. Time sequence A has global density
$\phi_{0}=-0.6$ (giving bubbles), while sequence B has $\phi_{0}=0.4$
giving droplets. (Sequences for $\phi_{0}=+0.6,-0.4$ can be generated
by inverting the color scale.) Both sequences have $\tilde{\kappa}=-0.1$.
Panel C shows the mean steady state droplet size $\bar{R}$ (defined
as $(A_{d}/4\pi)^{1/2}$ with $A_{d}$ the droplet area) on variation
of $\tilde{\kappa}$. The best linear fit on log-log gives exponent
$-0.52$, in good agreement with the scaling argument of \eqref{eq:Rscale}.
System size $256^{2}$; for movies and simulation details see \cite{siFields}.
\label{fig:1}}
\end{figure*}

{\em Active scalar field theory:} Our starting point is the diffusive
dynamics of a conserved scalar field $\phi(\boldsymbol{r},t)$ in
a momentum-conserving fluid of velocity $\boldsymbol{v}(\boldsymbol{r},t)$:
\begin{alignat}{1}
\dot{\phi} & +\boldsymbol{\nabla}\cdot\boldsymbol{J}+\boldsymbol{v}\cdot\boldsymbol{\nabla}\phi=0.\label{eq:orderPara}
\end{alignat}
Here $\boldsymbol{J}$ is the current density of $\phi$, which contains
equilibrium, active and stochastic contributions. Keeping active terms
to order $\mathcal{O}(\phi^{2},\nabla^{3})$ in \eqref{eq:orderPara},
$\boldsymbol{J}$ obeys \cite{wittkowski2014scalar,tiribocchi2015active,tjhung2018cluster}
\begin{subequations}\label{eq:orderPara-1} 
\begin{alignat}{1}
\boldsymbol{J} & =M\left(-\boldsymbol{\nabla}\mu+\zeta(\nabla^{2}\phi)\boldsymbol{\nabla}\phi\right)+\sqrt{2DM}\boldsymbol{\Lambda},\\
\mu & =\mu^{E}+\mu^{\lambda},\quad\mu^{E}=\frac{\delta\mathcal{F}}{\delta\phi},\quad\mu^{\lambda}=\lambda|\boldsymbol{\nabla}\phi|^{2}.
\end{alignat}
\end{subequations} Here $M$ is a mobility, assumed constant (we
set $M=1$ in what follows); $\boldsymbol{\Lambda}$ is a zero-mean,
unit-variance Gaussian white noise, and $D$ is a noise temperature
\cite{phiDependence}. The equilibrium and non-equilibrium parts of
the chemical potential for $\phi$ are denoted by $\mu^{E}$ and $\mu^{\lambda}$,
while $\mathcal{F}$ is the Landau-Ginzburg free energy functional:
$\mathcal{F}[\phi]=\int\left(\frac{a}{2}\phi^{2}+\frac{b}{4}\phi^{4}+\frac{\kappa}{2}(\boldsymbol{\nabla}\phi)^{2}\right)d\boldsymbol{r}$,
which gives bulk phase separation for $a<0$, with $b,\kappa>0$ for
stability \cite{bray1994theory,chaikin2000principles}.

The terms in $\zeta$ and $\lambda$ in \eqref{eq:orderPara-1} break
time-reversal symmetry at $\mathcal{O}(\phi^{2},\nabla^{4})$ in \eqref{eq:orderPara}.
These terms also break the $\phi\rightarrow-\phi$ symmetry of the
passive limit; however the full system of equations remains invariant
under $(\phi,\lambda,\zeta)\rightarrow-(\phi,\lambda,\zeta)$. This
means that all statements made below about droplets also apply to
the phase-inverted case of bubbles, so long as $\lambda$ and $\zeta$
are also changed in sign. Notably, the reverse Ostwald process stabilizes
{\em only} droplets for $\zeta<0$ and {\em only} bubbles for
$\zeta>0$ \cite{tjhung2018cluster}.

The fluid flow, in the limit of low Reynolds number (as applicable
to microswimmers), is obtained from the solution of the Stokes equation:
$\boldsymbol{\nabla}\cdot\boldsymbol{\sigma}=-\boldsymbol{f},$ where
$\boldsymbol{f}=\boldsymbol{\nabla}\cdot(\boldsymbol{\Sigma}^{A}+\boldsymbol{\Sigma}^{E})$
is the force density on the fluid, $\bm{\sigma}=-p\boldsymbol{I}+\eta(\bm{\nabla}\boldsymbol{v}+(\bm{\nabla}\boldsymbol{v})^{T})$
is the Cauchy fluid stress, $\eta$ is viscosity, $\boldsymbol{I}$
is the identity tensor, and $p$ is the pressure field which contains
all isotropic terms and ensures incompressibility ($\nabla\cdot\boldsymbol{v}=0$)
\cite{landau1959fluid}. We neglect noise in these flow equations
since this would involve the thermal temperature $T$ which is vastly
smaller than $D$ for active swimmers.

There exists a standard procedure \cite{kendon2001inertial} to derive
the deviatoric stress $\boldsymbol{\Sigma}^{E}$ in equilibrium systems
using the free energy $\mathcal{F}$. This stress, retaining all isotropic
terms, satisfies $\boldsymbol{\nabla}\cdot\boldsymbol{\Sigma}^{E}=-\phi\nabla\mu^{E}$,
which is the thermodynamic force density on the fluid due to gradients
of the concentration field $\phi$ \cite{bray1994theory}. The deviatoric
stresses $\boldsymbol{\Sigma}^{E}$ and $\boldsymbol{\Sigma}^{A}$
are then, in $d$-dimensions, given to the required order as 
\begin{gather}
\boldsymbol{\Sigma}^{E}=-\kappa\boldsymbol{S},\qquad\boldsymbol{\Sigma}^{A}=-(\tilde{\kappa}-\kappa)\boldsymbol{S},\label{eq:activeStress}
\end{gather}
where $\boldsymbol{S}\equiv(\boldsymbol{\nabla}\phi)(\boldsymbol{\nabla}\phi)-\tfrac{1}{d}|\boldsymbol{\nabla}\phi|^{2}\boldsymbol{I}$.
The mechanical stress $\boldsymbol{\Sigma}^{A}+\boldsymbol{\Sigma}^{E}=-\tilde{\kappa}\boldsymbol{S}$
is not derived from a free energy and breaks detailed balance in general.
Its overall coefficient $\tilde{\kappa}$ can be either positive (for
extensile microswimmers) or negative (for sufficiently contractile
ones) \cite{tiribocchi2015active}, unlike equilibrium systems where
$\boldsymbol{\Sigma}^{A}=0$.

Equations (\ref{eq:orderPara}-\ref{eq:activeStress}) define our
Active Model H for the diffusive dynamics of a conserved order parameter
with momentum conservation.The numerical method we use to integrate
these equations is described in \cite{siFields}. Having neglected
inertia, they reduce to an effective dynamics for $\phi$ alone, which
reads (with $M=1$) 
\begin{equation}
\dot{\phi}=\nabla^{2}\mu^{\text{eff}}-\boldsymbol{\nabla}\phi(\boldsymbol{r})\cdot\int\mathbf{G}(\boldsymbol{r}-\boldsymbol{r}')\cdot\boldsymbol{f}(\boldsymbol{r}')\,d\boldsymbol{r}'-\boldsymbol{\nabla}\cdot\boldsymbol{J}^{\Lambda}.\label{eq:effective}
\end{equation}
Here $\mathbf{G}(\boldsymbol{r})$ is the Oseen tensor, $\boldsymbol{J}^{\Lambda}=\sqrt{2D}\boldsymbol{\Lambda}$,
and $\mu^{\text{eff}}=\mu^{E}+\mu^{\lambda}+\mu^{\zeta}$ an effective
chemical potential. This has a part $\mu^{\zeta}$, constructed via
Helmholtz decomposition of the active current term in \eqref{eq:orderPara-1},
as $\zeta(\nabla^{2}\phi)\boldsymbol{\nabla}\phi=-\boldsymbol{\nabla}\mu^{\zeta}+\boldsymbol{\nabla}\times\boldsymbol{A}$.
The $\boldsymbol{\nabla}\times\boldsymbol{A}$ term is divergenceless,
and so cannot contribute to $\dot{\phi}$ in \eqref{eq:effective},
allowing it to be ignored \cite{tjhung2018cluster}.

{\em Three tensions:} At large scales, the dynamics of an interface
between phases is controlled by its curvature and its interfacial
tension. Without activity, there is only one such tension, $\gamma_{{0}}$$=\sqrt{-8\kappa a^{3}/9b^{2}}$,
governing both diffusive and mechanical sectors; curvature then drives
diffusive currents and/or fluid flow via Laplace pressure \cite{bray1994theory,cates2018theories}.

With activity, two further tensions enter \cite{tjhung2018cluster,tiribocchi2015active}.
First, the chemical potential term $\mu^{\zeta}$ is in general nonlocal,
and acquires a step-discontinuity across a curved interface that is
cancelled by a counter-step in $\mu^{E}$. The result is a discontinuity
in $\phi$ different from the equilibrium one. This is captured by
a `pseudotension' $\gamma_{\phi}$ that replaces $\gamma_{{0}}$ when
calculating diffusive fluxes between droplets, and becomes negative
for sufficiently negative $\zeta,\lambda$. Crucially, negative $\gamma_{\phi}$
does not make interfaces locally unstable; spherical droplets stay
spherical. Rather, its effect is to drive the system towards states
of globally uniform curvature (monodisperse drops) whereas positive
tension promotes curvature differences by shrinking small droplets
and growing large ones (the Ostwald process) \cite{tjhung2018cluster}.

The third tension arises from the active stress at interfaces, where
swimmers align parallel or antiparallel to the surface normal. In
either case, contractile swimmers pull fluid inward along the normal
direction and expel it in the interfacial plane, causing stretching,
whereas extensile swimmers do the opposite. The resulting mechanical
tension is $\gamma_{v}=\tilde{\kappa}\gamma_{{0}}/{\kappa}$, which
is negative for sufficient contractility; this is known to interrupt
phase separation for bicontinuous regimes \cite{tiribocchi2015active}.
We find next that it also does so for droplets, by destabilizing interfaces
{\em locally} (in contrast with negative $\gamma_{\phi}$; see
above).

\begin{figure}[b]
\centering\includegraphics[width=0.44\textwidth]{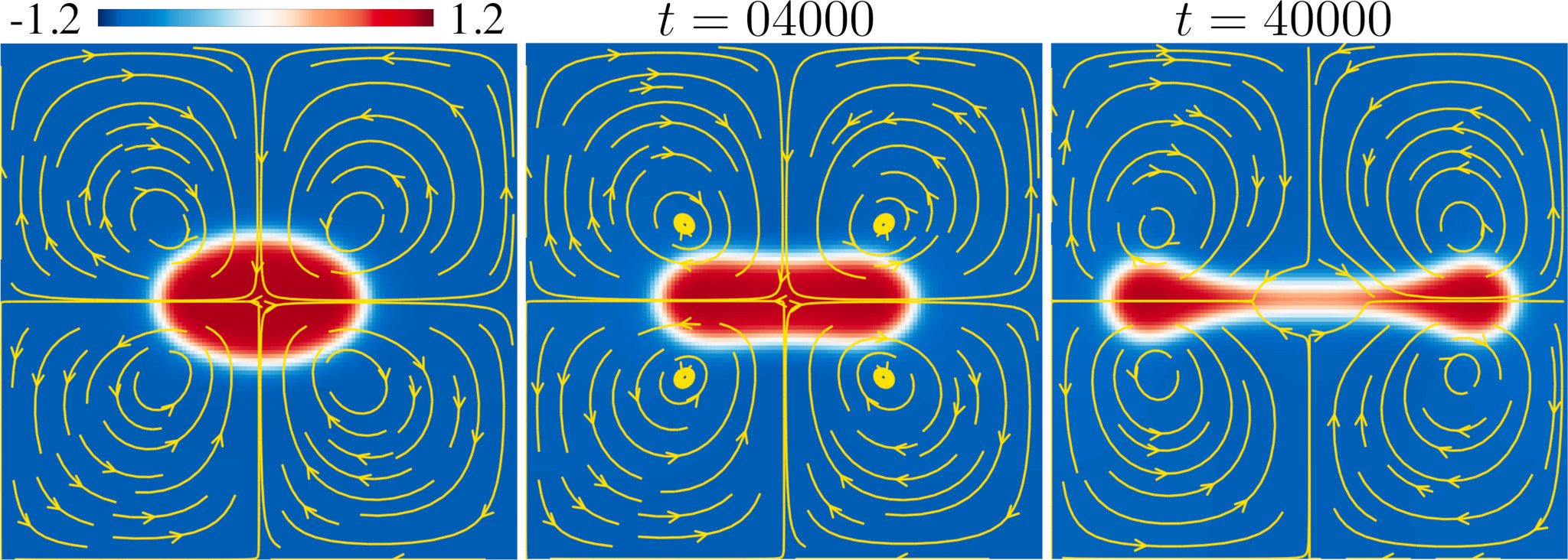}\caption{Snapshots mapping the order parameter and flow streamlines, starting
from a deformed droplet for a sufficiently contractile active stress
(self-shearing instability).\label{fig:2}}
\end{figure}

{\em Self-shearing instability:} For simplicity we first consider
active mechanical stress alone, setting $\lambda=\zeta=0$ so that
$\gamma_{\phi}=\gamma_{{0}}>0$ while $\tilde{\kappa}=-0.1$, giving
negative $\gamma_{v}$. In Fig.(\ref{fig:1}), we show the dynamic
interruption of phase separation in two dimensions and the resulting
steady state droplet size $\bar{R}$. The negative mechanical tension,
arising from contractile stress, results in a self-shearing of the
droplets, causing large ones to split. This is balanced by Ostwald
ripening: small droplets evaporate while large ones grow until they
in turn become unstable. The result is a dynamical steady state of
droplet splitting followed, on average, by diffusive growth of one
offspring and disappearance of the other(s). Supplemental movies show
this dynamics clearly \cite{siFields}. In Fig.(\ref{fig:2}), and
supplemental movie \cite{siFields}, we show the dynamics of an individual
droplet for negative $\gamma_{v}$, which exhibits the flow-induced
droplet breakup mechanism. Our numerics do not of course directly
introduce a negative interfacial tension, but solve the full equations
(\ref{eq:orderPara}-\ref{eq:activeStress}) or equivalently \eqref{eq:effective}.

{\em Scaling of droplet size:} In Fig.\ref{fig:1}C, we show that
the droplet size obeys $\bar{R}\sim|\gamma_{v}/\gamma_{{\phi}}|^{-0.52}$
(best fit exponent), where, in these simulations, $\gamma_{\phi}=\gamma_{{0}}>0$.
We now argue on simple grounds for a negative one half exponent whenever
$\gamma_{v}<0$ and $\gamma_{\phi}>0$. From the mechanical tension
$\gamma_{v}$ and fluid parameters we can construct just one quantity
with the dimensions of velocity: ${\cal V}_{v}=\gamma_{v}/\eta$.
This is the familiar coarsening rate $\dot{L}$ for systems with bicontinuous
domains of size $L$, in the so-called `viscous hydrodynamic' regime
where curvature drives fluid motion and diffusive fluxes are negligible
\cite{kendon2001inertial}. For a droplet of size $\bar{R}$ with
negative $\gamma_{v}$, one thus expects the time between scission
events to scale as $\tau=-\bar{R}/{\cal V}_{v}$. The Ostwald process
gives another speed, which is the rate of change of the mean droplet
size ${\cal V}_{\phi}(\bar{R})=\dot{\bar{R}}\propto{M\gamma_{\phi}}/{\phi_{B}^{2}\bar{R}^{2}}$
where $\phi_{B}$ is the binodal density \cite{bray1994theory,tjhung2018cluster}.
Balancing a positive Ostwald speed ${\cal V}_{\phi}(\bar{R})$ with
a negative hydrodynamic speed ${\cal V}_{v}$ gives the promised scaling
law 
\begin{equation}
\bar{R}\propto\left(\frac{-\gamma_{v}\phi_{B}^{2}}{\eta M\gamma_{\phi}}\right)^{-1/2}\sim\left|\frac{\gamma_{v}}{\gamma_{\phi}}\right|^{-1/2}\,.\label{eq:Rscale}
\end{equation}

The predicted one-half exponent is in excellent agreement with Fig.\ref{fig:1}C.
Unlike the simulations, our scaling argument does not include noise,
which appears inessential to the steady state, at least in the parameter
range investigated here. We checked this by explicitly switching off
noise once the steady state is achieved and found that it continues
to exist, with essentially the same droplet life-cycle and mean size.
While noise is needed initially to start the phase separation, once
enough droplets are present the contractile active stress can maintain
the dynamics of scission and coarsening indefinitely. In addition
to splitting and ripening, we observe splitting-induced coalescence
mediated by fluid flow. This resembles the coalescence-induced coalescence
mechanism seen for semidilute passive droplets in two dimensions \cite{wagner2001phase}.
However, this cannot change the scaling because its physics is also
governed by ${\cal V}_{\phi}$.

\begin{figure}[t]
\centering\includegraphics[width=0.475\textwidth]{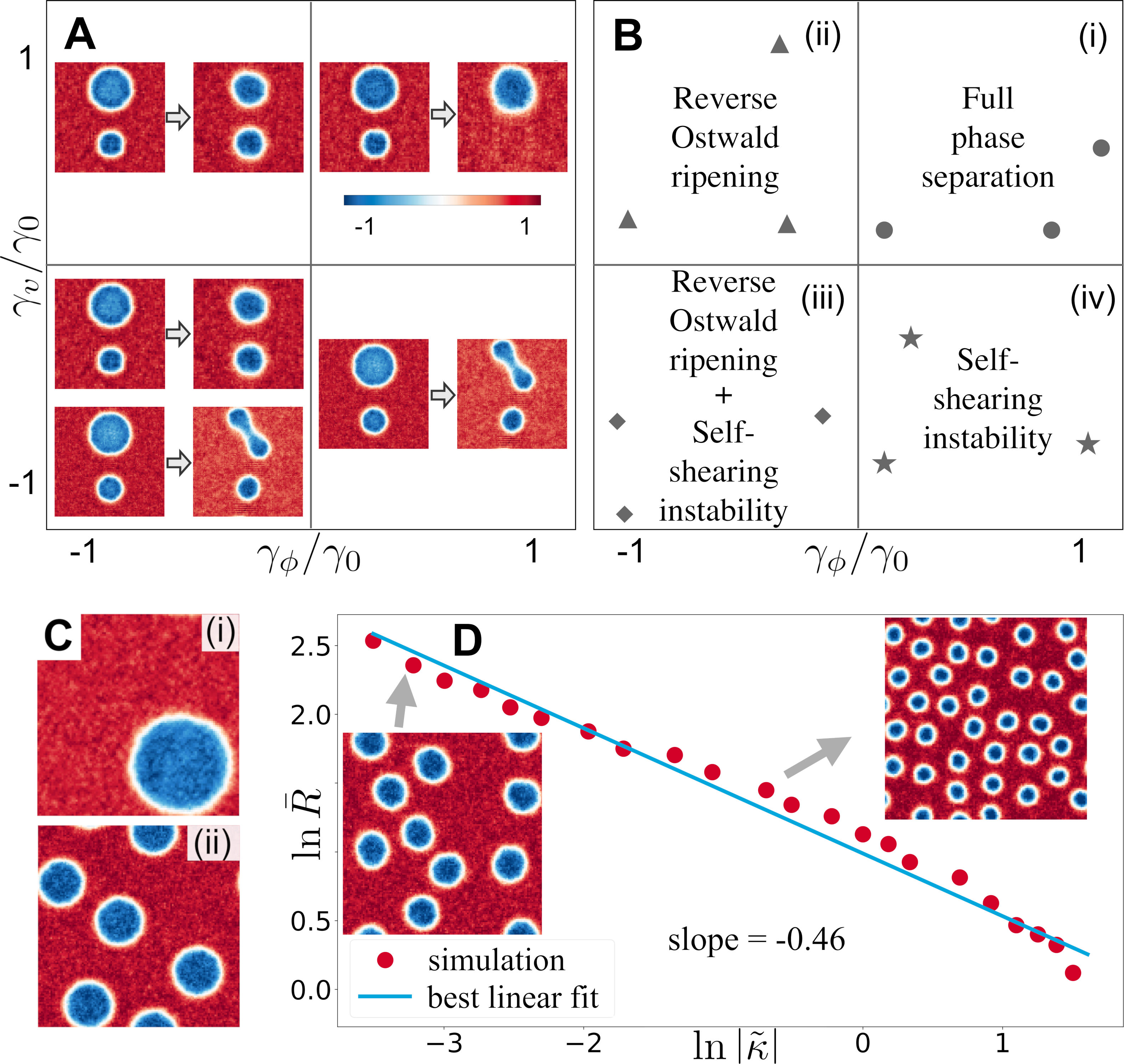}\caption{Phases of Active Model H in the plane $(\gamma_{\phi},\gamma_{v})$.
Panel A shows the dynamics of two unequal size bubbles in four regions
distinguished by the signs of $\gamma_{\phi}$ and $\gamma_{v}$,
as labelled in B. The markers denote simulation points. Snapshots
of steady states are shown in panel C(i-ii) for the regions in panel
B(i-ii). Panel D shows that $\bar{R}\sim|\gamma_{v}|^{-1/2}$ (\ref{eq:Rscale})
continues to hold (best fit exponent is $-0.46$) in the region B(iii).
See Fig(\ref{fig:1}) for steady states of B(iv). System size is $128^{2}$.
\label{fig:3}}
\end{figure}

{\em Competing activity channels}: We next consider the full dynamics
of (\ref{eq:effective}), where both the active tensions can change
sign, and enumerate the possible steady-states. The resulting phase
diagram, in the $(\gamma_{\phi},\gamma_{v})$ plane, is shown in Fig.(\ref{fig:3}A-B).
In panel A, we show time evolution of two bubbles of unequal sizes
and their final steady state. The two bubbles disproportionate if
both tensions are positive; the corresponding outcome in the many-droplet
system (panel B) is forward Ostwald dynamics. The three activity parameters
$\tilde{\kappa},\lambda,\zeta$ in this region serve merely to renormalize
the passive behavior. In contrast, when $\gamma_{v}>0$ and $\gamma_{\phi}<0$,
we recover the result of \cite{tjhung2018cluster}, with reverse Ostwald
ripening arresting phase separation of droplets (see panel C(ii)).
As in the forward case, the reverse Ostwald regime entails little
or no fluid motion, so the results given in \cite{tjhung2018cluster}
apply even for the wet systems studied here, with physics controlled
by $\gamma_{\phi}$ and no role for $\gamma_{v}$. Thirdly, the case
$\gamma_{v}<0$ and $\gamma_{\phi}>0$ is governed by our interrupted
steady state with the splitting/coarsening life-cycle described above;
here $\lambda$ and $\zeta$ enter only via $\gamma_{\phi}$. Because
the steady state balances mechanical and diffusive processes, both
tensions enter the mean droplet size via \eqref{eq:Rscale}.

In the final regime, both tensions are negative. Here, because negative
$\gamma_{\phi}$ reverses Ostwald ripening, the balance of splitting
and ripening embodied in \eqref{eq:Rscale} cannot be maintained.
Were splitting to continue, the number of droplets would increase
forever, with the reverse Ostwald process driving these towards uniformity
in size but not allowing the number to reduce by evaporation. Accordingly,
for negative $\gamma_{\phi}$, no matter how small its magnitude,
the steady state {\em must} consist of almost static droplets with
no splitting, and this is indeed what we observe.

Despite this complete change of mechanism, we now argue that the steady-state
droplet size is still governed by $\bar{R}\sim|\gamma_{v}|^{-1/2}$
as in \eqref{eq:Rscale}. This is because the reverse Ostwald process
drives the system towards uniformity of curvature which, for a single
droplet, directly opposes the self-shearing instability. Indeed, for
an amplitude ${\cal A}\sim\epsilon R$ of the lowest deformation mode
in a droplet of radius $R$, one expects on dimensional grounds that,
up to prefactors, $\dot{{\cal A}}\sim\epsilon({\cal V}_{v}+{\cal V}_{\phi}(R))$.
Here the first term is the self-shearing instability and the second
is stabilizing for negative $\gamma_{\phi}$. Stability is restored
for droplets smaller than $\bar{R}$ given by \eqref{eq:Rscale},
which thus again gives the scaling of steady-state droplet size, albeit
by a completely different mechanism of arrest rather than interruption.
This scaling is consistent with simulations, see Fig.(\ref{fig:3}D).

All the above arguments apply equally to bubbles as to droplets via
invariance under $(\phi,\lambda,\zeta)\to-(\phi,\lambda,\zeta)$.

{\em Conclusion:} In wet active systems containing droplets or
bubbles, microphase separation can replace bulk phase separation by
three distinct mechanisms. These are reverse Ostwald ripening (also
present in the dry limit, and caused by negative pseudotension $\gamma_{\phi}$);
hydrodynamic interruption by self-shearing (caused by negative mechanical
tension $\gamma_{v}$); and a peculiar combination of the two when
both of tensions are negative. In contrast to reverse Ostwald ripening,
where the size of droplets or bubbles in the arrested steady state
is selected by noise (or, in its absence, initial conditions) \cite{tjhung2018cluster},
our self-shearing mechanism continues to operate even if the noise
is switched off after the steady-state is reached. The droplet size
is then selected by a balance between droplet splitting and forward
Ostwald ripening. This balance is unsustainable when $\gamma_{\phi}$
is {\em also} negative, and so is replaced by a new one whereby
the reverse Ostwald process suppresses the self-shearing instability
for small droplets, leading to static arrest rather than dynamic interruption.

These results point to a subtle interplay of causes behind the phenomena
of active microphase separation. We contend that they cannot be understood
without first clearly enumerating the relevant activity channels,
and then studying their interaction; this is best done within the
minimalist framework of an active field theory as done here. A possible
generalization of \eqref{eq:effective} would allow for boundary conditions
imposed on the fluid flow via a nearby wall; these are known to play
a crucial role in active phase separation \cite{singh2016crystallization,thutupalli2018FIPS,singh2019competing}
but with no clear understanding, so far, of the microphase-separated
case.

Our findings should complement more detailed work that could connect
our activity channels to microscopic interactions \cite{liebchen2015clustering,liebchen2017phoretic,alarcon2017morphology,sahaPRE2014,agudoPSPRL2019}.
They may also be be relevant to continuum models of multiple species
(some non-conserved \cite{zwicker2017growth}) that were used recently
to study microphase separations used by cells to create cytoplasmic
and nucleoplasmic organization \cite{shin2017liquid,weber2019physics}.
Fluid motion is often neglected in such studies but we have shown
that it brings new features (a similar case of elasticity in polymer
networks has been shown to arrest phase separation \cite{style2018liquid}),
suggesting exciting directions for future work.

\emph{Acknowledgements}: We thank Ronojoy Adhikari and Elsen Tjhung
for useful discussions. RS is funded by a Royal Society-SERB Newton
International Fellowship. MEC is funded by the Royal Society. Numerical
work was performed on the Fawcett HPC system at the Centre for Mathematical
Sciences. Work funded in part by the European Research Council under
the Horizon 2020 Programme, ERC grant agreement number 740269.

\appendix
\vspace{.4cm}

\section*{Supplemental information (SI)}

\section{Methodology}

In this section, we explain the numerical methods used to study the
field theory of an active scalar field with mass and momentum conservation.
We first explain the solution of the flow equations and then given
details of the simulation and provide the system of parameters. 
\begin{table*}
\global\long\def\arraystretch{2.2}

\begin{tabular}{|>{\centering}b{2cm}|>{\centering}p{3cm}|>{\centering}p{3cm}|>{\centering}p{2cm}|>{\centering}p{2cm}|>{\centering}p{2cm}|}
\hline 
Figure  & Global density $\phi_{0}$  & System size  & $\tilde{\kappa}$  & $\lambda$  & $\zeta$\tabularnewline
\hline 
\hline 
1 A  & 0.6  & $256\times256$  & -0.1  & 0  & 0\tabularnewline
\hline 
1 B  & -0.4  & $256\times256$  & -0.1  & 0  & 0\tabularnewline
\hline 
1 C  & 0.6  & $128\times128$  & varied  & 0  & 0\tabularnewline
\hline 
2  & -0.1  & $128\times128$  & -0.1  & 0  & 0\tabularnewline
\hline 
3 A  & 0.6  & $128\times128$  & varied  & varied  & varied\tabularnewline
\hline 
3 B  & 0.6  & $128\times128$  & varied  & varied  & varied\tabularnewline
\hline 
3 C(i)  & 0.6  & $128\times128$  & 1.1  & 0  & 0\tabularnewline
\hline 
3 C(ii)  & 0.6  & $128\times128$  & 1.1  & 1.2  & 2\tabularnewline
\hline 
3 D  & 0.6  & $128\times128$  & varied  & 1.2  & 2\tabularnewline
\hline 
\end{tabular}\caption{\label{tab:Table-of-simulation-parameters}Simulation parameters used
to study the field theory of an active scalar with mass and momentum
conservation. Throughout the paper, the following parameters have
been kept fixed $a=-0.25$, $b=0.25$, $\kappa=1$, $M=1$, and $D=0.02$.
The value of $\tilde{\kappa}$ has been varied in Fig.(1C) and Fig.(3D)
to show the scaling of the length scale with contractile activity.
We start with an elliptic droplet in 2A and 2B. The parameters for
activity in mass conservation ($\lambda$, $\zeta$) and momentum
conservation ($\tilde{\kappa}$) equations has been varied in Fig.(3A-B)
to enumerate distinct steady states. All the simulations reported
are in two space dimensions, while the predictions of the manuscript
would be maintained in higher dimensions.}
\end{table*}

\subsection{Stokes solver}

At low Reynolds number, the fluid flow satisfies the Stokes equation
\begin{alignat}{1}
-\boldsymbol{\nabla}p+\eta\nabla^{2}\boldsymbol{v} & =-\boldsymbol{f},\label{eq:stokes1}\\
\boldsymbol{\nabla}\cdot\boldsymbol{v} & =0.
\end{alignat}
Here, 
\[
\boldsymbol{f}=\boldsymbol{\nabla}\cdot(\boldsymbol{\Sigma}^{A}+\boldsymbol{\Sigma}^{E}),
\]
is the force density on the fluid. We now show how to determine flow
given the force density and detail the implementation of incompressibility
in the flow. We use Fourier transforms to obtain a solution of the
above equations. We define the Fourier transform of a function $\varphi(\boldsymbol{r})$
as\begin{subequations}\label{eq:FT} 
\begin{gather}
\hat{\varphi}(\boldsymbol{k})=\mathbb{\mathbb{F}}\left[\varphi(\boldsymbol{r})\right]=\int\varphi(\boldsymbol{r})\,e^{-i\boldsymbol{k}\cdot\boldsymbol{r}}\,d\boldsymbol{r},\\
\varphi(\boldsymbol{r})=\mathbb{\mathbb{F}}^{-1}\left[\hat{\varphi}(\boldsymbol{k})\right]=\frac{1}{(2\pi)^{3}}\int\hat{\varphi}(\boldsymbol{k})\,e^{i\boldsymbol{k}\cdot\boldsymbol{r}}\,d\boldsymbol{k}.
\end{gather}
\end{subequations}We now Fourier transform (\ref{eq:stokes1}) to
obtain\begin{subequations} 
\begin{align}
-i\boldsymbol{k}\hat{p}-\eta k^{2}\boldsymbol{\hat{v}} & =-\boldsymbol{\hat{f}},\label{eq:stokes-k}\\
i\boldsymbol{k}\cdot\boldsymbol{\hat{v}} & =0.\label{eq:incompressibility}
\end{align}
\end{subequations}The above equations can then be used to obtain
the Fourier transform of the pressure field 
\begin{equation}
\hat{p}=-i\boldsymbol{k}\cdot\boldsymbol{\hat{f}}/k^{2}.
\end{equation}
The above expression of the pressure is then used in (\ref{eq:stokes-k})
to obtain the solution of the fluid flow, with built-in incompressibility,
given as 
\begin{align}
\boldsymbol{\hat{v}} & =\mathbf{\hat{G}}(\boldsymbol{k})\cdot\boldsymbol{\hat{f}},\quad\quad\mathbf{\hat{G}}(\boldsymbol{k})=\frac{1}{\eta}\left(\frac{\boldsymbol{I}}{k^{2}}-\frac{\boldsymbol{k}\boldsymbol{k}}{k^{4}}\right).\label{eq:stokes-k-1}
\end{align}
Here $\mathbf{\hat{G}}(\boldsymbol{k})$ is the Fourier transform
of the Oseen tensor $\mathbf{G}(\boldsymbol{r})$ \cite{pozrikidis1992}.
The explicit forms of the Oseen tensor, in three dimensions, is 
\begin{alignat}{1}
\mathbf{G}(\boldsymbol{r}) & =\frac{1}{8\pi\eta}\left(\frac{\boldsymbol{I}}{r}+\frac{\boldsymbol{r}\boldsymbol{r}}{r^{3}}\right).
\end{alignat}
The above is the Green's function of Stokes equation in an unbounded
domain and captures the long range interactions at low Reynolds number
\cite{happel1965low}. It should be noted that we are confining ourselves
to low Reynolds number by not solving the full Navier-Stokes equation.
Thus, our theory is mainly applicable to active matter systems of
microorganisms \cite{lauga2009} and synthetic microwswimmers \cite{ebbens2010pursuit},
where the Reynolds number is known to be very small.

\subsection{Simulation details}

The simulations reported in this manuscript are performed by the numerical
integration of Eq.(1-3) of the main text. Eq.(\ref{eq:stokes-k-1})
is the solution for the fluid flow, which we use for numerical computations.
The solution, by construction, satisfies incompressibility exactly.
The Fourier solution of the fluid flow is then used in Eq.(\ref{eq:orderPara})
to numerical update the order parameter field. The terms in the mass
conservations equations are computed using a pseudospectral method
(involving Fourier transforms and 2/3 dealiasing procedure) \cite{orszag1971elimination,boyd2001chebyshev}.
The spectral solver for the fluid flow and order parameter is numerically
implemented using standard fast Fourier transforms (FFTs) and by virtue
of these Fourier transforms, periodic boundary conditions are automatically
ensured in the system. We evolve the system in time using the explicit
Euler\textendash Maruyama method \cite{kloeden1992higher}. We provide
the parameters used in generating all the figures of the manuscript
in Table \ref{tab:Table-of-simulation-parameters}.

\section{Higher-order terms of active stress}

In the main text, we have provided a form of the active stress tensor
$\boldsymbol{\Sigma}^{A}$ in Eq.(3). This term preserves the $\phi\rightarrow-\phi$
symmetry of the passive model H. It is possible to add higher-order
terms of the form $\sim-\left(\alpha\phi\right)\,\boldsymbol{S}$
and $\sim-\left(\beta\nabla^{2}\phi\right)\,\boldsymbol{S}$ to the
active stress tensor. These terms break the $\phi\rightarrow-\phi$
and can be derived by taking into account the fact that self-propulsion
speed of active particles is dependent on their local density \cite{tailleur2008statistical,cates2015,tiribocchi2015active}.
In what follows, we show that such terms can never reverse the sign
of the active stress. In other words, it is not possible to make a
contractile swimmer by slowing down an extensile swimmer. 

To obtain terms of the form $\sim-\left(\alpha\phi\right)\,\boldsymbol{S}$
and $\sim-\left(\beta\nabla^{2}\phi\right)\,\boldsymbol{S}$, we note
that the density-dependent speed $w([\phi])$ of active particles
can be written as $w([\phi])=w_{0}+w_{1}\phi+w_{2}\nabla^{2}\phi+\text{higher order terms}$
\cite{tailleur2008statistical}. The active stress can be written
as $\boldsymbol{\Sigma}^{A}\sim\xi_{P}p_{i}p_{j}$, where $p_{i}=-\tau\nabla_{i}(w\phi)$
and $\tau$ is the orientational relaxation time \cite{tiribocchi2015active}.
Thus, it follows that the active stress is $\Sigma_{ij}^{A}=\Upsilon S_{ij}$,
with $\Upsilon\sim\xi_{P}\tau^{2}(w_{0}+2w_{1}\phi+w_{2}\nabla^{2}\phi+\text{h.o.t})^{2}$.
It is then clear that the sign of the coefficients in the active stress
is determined by the parameter $\xi_{P}$, which is positive for extensile
swimmer and negative for contractile swimmers. For simplicity, we
have treated $\xi_{P}$ as constant but making it a function of $w$
has no effect on our conclusions in the paper. Moreover, by construction,
the terms proportional to $\alpha$ and $\beta$ can only reduce the
magnitude but never reverse the sign of the active stress. Thus, in
this paper, we consider a constant $\Upsilon$ which captures all
the qualitative effects.

\section{Supplemental movies}

In this Appendix, we describe the supplemental movies which are complementary
with the main text. These movies are available online at \href{https://www.dropbox.com/sh/5krtwg7gwmq736r/AADZMIimLTgjVdyfb1EYnJZGa?dl=0}{at this URL.}

\textbf{Movie I: }The movie shows the dynamics of Fig.1A in the main
text, which is the self-shearing instability in the bubble phase for
contractile active stress ($\tilde{\kappa}<0$) with no activity in
the diffusive sector. Starting from a uniform phase, droplets nucleate
due to the noise and they grow in size until interrupted by the contractile
active stress. The parameters are $a=-0.25$, $b=0.25$, $\kappa=1$,
$\tilde{\kappa}=-0.1$, $\lambda=0$, and $\zeta=0$. The initial
global density $\phi_{0}=0.6$ and the system size is $256\times256$.

\textbf{Movie II: }Same as in movie I but with $\phi_{0}=-0.4$. It
corresponds to Fig.(1B) of the main text.

\textbf{Movie III: }The movie shows the fluid flow and dynamics of
deformed droplet in the presence of a sufficiently contractile active
stress (self-shearing instability). It corresponds to Fig(2) of the
main text. The parameters are $a=-0.25$, $b=0.25$, $\kappa=1$,
$\lambda=0$, $\zeta=0$, and $\tilde{\kappa}=-0.1$. The system size
used for this simulation is $256\times256$. 

\textbf{Movie IV: }Nucleation and growth of bubbles when activity
is present in both the mechanical and diffusive sectors. A snapshot
from the steady state is in Fig(3C-(ii)) of the main text. The parameters
are $a=-0.25$, $b=0.25$, $\kappa=1$, $\tilde{\kappa}=1.1$, $\lambda=1$,
$\zeta=2$. The initial global density $\phi_{0}=0.6$ and the system
size is $128\times128$.

\textbf{Movie V: }Same as in movie IV, but with $\tilde{\kappa}=-0.1$.
It corresponds to Fig.(3D) in the main text. The fluid due to contractile
stress has the effect of stretching the interface, which is the precursor
to the self-shearing instability as the system approaches the steady
state. It should be noted that there is no self-shearing in the steady
state - this is because the reverse Ostwald process drives the system
towards uniformity of curvature which directly opposes the self-shearing
instability.
\end{document}